\documentclass[preprint,aps]{revtex4}
\usepackage{amsmath}
\usepackage{graphicx}

\begin{document}

\title{ \emph{ T. Dauxois' ``Non-Gaussian distributions under scrutiny"} under scrutiny \footnote{To appear in {\it Astrophysics and Space Science} (Springer, 2008), in press.}}

\author{Constantino Tsallis}%
\email{tsallis@cbpf.br}

\affiliation{Centro Brasileiro de Pesquisas Fisicas \\ 
Rua Xavier Sigaud 150, 22290-180 Rio de Janeiro-RJ, Brazil\\
and\\
Santa Fe Institute, 1399 Hyde Park Road, Santa Fe, NM 87501, USA}

\begin{abstract}
A recent paper by T. Dauxois entitled ``Non-Gaussian distributions under scrutiny" is submitted to scrutiny. Several comments on its content are made, which constitute, at the same time, a brief state-of-the-art  review of nonextensive statistical mechanics, a current generalization of the Boltzmann-Gibbs theory. Some inadvertences and misleading sentences are pointed out as well. 
\end{abstract}


\maketitle

T. Dauxois has recently commented \cite{Dauxois2007} an interesting paper by Hilhorst and Schehr \cite{HilhorstSchehr2007}. In the latter paper, two specific probabilistic models, namely a discrete one \cite{MoyanoTsallisGellMann2006} (from now on referred to as the {\it MTG model}) and a continuous one \cite{ThistletonMarshNelsonTsallis2006} (from now on referred to as the {\it TMNT model}), are {\it analytically} discussed. Both models consist in $N$ correlated random variables (respectively discrete and continuous), and the point under study is what is the limiting distribution when $N \to\infty$.  {\it Numerical indications}$^{1}$\footnotetext[1]{See the title of the present Ref. [3].} 
have been found that quite strongly suggest that these limiting distributions could be of the $q$-Gaussian class (Eq. (1) in \cite{Dauxois2007}). Hilhorst and Schehr have shown \cite{HilhorstSchehr2007} that they are {\it not}, even if they are numerically intriguingly close. The relations between the entropic indices $q$ (of the possible $q$-Gaussian limiting distributions) and the parameters of the models have been addressed as well. It was analytically confirmed in \cite{HilhorstSchehr2007} the correctness of the relations conjectured respectively in  \cite{MoyanoTsallisGellMann2006} and in \cite{ThistletonMarshNelsonTsallis2006}, if these {\it analytically exact} non $q$-Gaussian distributions were to be approached by $q$-Gaussians!

Why may all this have some general interest? The reason lies on the fact that $q$-Gaussian distributions play a special role in nonextensive statistical mechanics \cite{Tsallis1988,CuradoTsallis1991,TsallisMendesPlastino1998}, a current generalization (based on the entropy (2) of \cite{Dauxois2007}) of the celebrated BG theory. Recent reviews of $q$-statistics and a vast set of experimental, observational and computational applications and/or verifications can be seen in \cite{GellMannTsallis2004,BoonTsallis2005,AbeHerrmannQuaratiRapisardaTsallis2007,Tsallis2007,biblio}. Very specifically, it has been recently {\it proved} \cite{UmarovTsallisSteinberg2006} a $q$-generalization of the Central Limit Theorem for $q \ge 1$. More precisely, if we have $N$ random variables that are strongly correlated in a special manner (called {\it $q$-independence}; see \cite{UmarovTsallisSteinberg2006} for details), it can be rigorously proved that their sum approaches, when appropriately centered and scaled, a $q$-Gaussian distribution. The standard CLT is recovered as the $q=1$ particular instance. The proof is under progress for $q<1$ (which are distributions with a compact support, even in the $N \to\infty$ limit). Nevertheless, preliminary studies suggest that the  $q$-CLT should remain applicable even for $q<1$. Since the {\it MTG} and {\it TMNT} models have a compact support (i.e., of the $q<1$ type), one expects $q$-Gaussians whenever the correlation is of the $q$-independent type. Given the results in  \cite{HilhorstSchehr2007}, one is naturally led to argue that the strong correlations in those two models are {\it not $q$-independent but only almost $q$-independent}. An interesting question remains then open. What physical ingredient have the {\it MTG} and {\it TMNT} models failed to incorporate? In other words, there is something which is still missing in those models in order for the correlation to be {\it exactly} $q$- independent. What is it? Progress is presently being achieved along this line (deeply related to asymptotic scale-invariance), but this remains outside the present scope, which primarily is the careful scrutiny of the paper  \cite{Dauxois2007}. Let us address now some of the weaknesses of that paper. \\

(i) In contrast with what one reads in \cite{Dauxois2007}, the $TMNT$ model is {\it not} at all described nor presented in its Ref. [3]. In fact, it is  not even mentioned there, and has never been published! $^{2}$\footnotetext[2]{The details of the $TMNT$ model, as well as all our numerical and graphical results, were transmitted privately by me to Hilhorst, who had heard, in Natal-Brazil in March 2007, an oral presentation of mine including that subject}.  
Dauxois visibly confuses with \cite{ThistletonMarshNelsonTsallis2007}, whose content {\it has absolutely nothing to do} with the model discussed by Hilhorst and  Schehr. Had the Author of  \cite{Dauxois2007} paid more attention to his own Ref. [3], or at least to its title, the confusion would have been avoided. \\

(ii) The statement {\it ``could be the basis for a generalized central limit theorem"} (below Eq. (1) of \cite{Dauxois2007}) reveals that the Author is possibly unaware of the proof existing since already some time in  \cite{UmarovTsallisSteinberg2006} (further extensions are possible along the lines of \cite{UmarovTsallisGellMannSteinberg2006,UmarovTsallis2007}). \\

(iii) The statement {\it ``The basis for this suggestion ..."} (above Eq. (2) of \cite{Dauxois2007}) reveals that the Author is unaware that the crucial point is {\it not} the isolated fact that $q$-Gaussians optimize $S_q$ (something which, in contrast with what is referred in \cite{Dauxois2007}, is {\it not even vaguely mentioned} in Ref. [2] of \cite{Dauxois2007}!), but rather the remarkable fact that $q$-Gaussians also happen to be \cite{PlastinoPlastino1995,TsallisBukman1996} {\it exact} stable solutions of the nonlinear Fokker-Planck equation (since many decades called {\it Porous Medium Equation}). The mere fact that $q$-Gaussians would optimize some specific entropic functional would certainly not be of any particular significance. The entire rationale of what was at the time a conjecture (and is now a proved theorem) can be seen in \cite{Tsallis2005}. \\

(iv) The qualification {\it ``in the absence of firm grounds..."} reveals that the Author of  \cite{Dauxois2007} is possibly unaware of the numerous rigorous results which precisely provide a firm mathematical basis for the entropy $S_q$ and the associated nonextensive statistical mechanics. Among many others, let us mention the $q$-generalization, respectively in \cite{Santos1997} and in \cite{Abe2000}, of Shannon's and of Khinchine's uniqueness theorems, the already mentioned $q$-central limit theorems in \cite{UmarovTsallisSteinberg2006}, the Lesche-stability \cite{Abe2002} and the Topsoe-factorizability \cite{Topsoe2006} of the entropy $S_q$, the analytical results in \cite{AnteneodoTsallis2003} for Langevin-like stochastic equations in the presence of multiplicative noise, the analytical results of \cite{BaldovinRobledo2002a,BaldovinRobledo2002b,BaldovinRobledo2004,MayoralRobledo2005} concerning unimodal dissipative one-dimensional maps, the analytical connection \cite{WilkWlodarczyk2000,Beck2001,BeckCohen2003} to the Beck-Cohen superstatistics, etc.  \\

(v) The Author of \cite{Dauxois2007} wonders whether the whole idea of $q$-statistics could be {\it ``just a nice idea and a powerful fitting function"}. Fitting functions typically have one or more fitting parameters. Let us focus on this point. Fitting parameters can be of quite different natures. The first, and in some sense more fundamental, class is constituted by those numbers for which no theory exists in contemporary physics that could produce them from more basic principles. In this class we have all the so-called {\it universal constants} of contemporary physics ($c$, $h$, $G$, and $k_B$), which, as well known, might not even be constants! They are obtained, nowadays with great precision, from fittings concerning basic physical laws such as the Planck law for the black-body radiation. A second class is constituted by those parameters that can ideally be deduced from first principles through some specific way involving some specific information. But, in many cases, the procedure turns out to be untractable, either because of mathematical difficulties, or because some of the necessary information is missing. Such is the case of the orbit of say Mars. Newtonian mechanics can in principle predict this orbit given, at some time, the positions and velocities of all masses of the planetary system. Since this information is unavailable (and even if it was available, a colossal computer would be needed to perform the calculations!), astronomers use, as a first approximation, the elliptic form of a Keplerian orbit and just fit this form (which can be easily obtained within Newtonian mechanics) in order to obtain the concrete values of the two axis of the ellipse. The entropic index $q$ belongs to this class. It can be in principle deduced from the microscopic dynamics of the system. And this is indeed done successfully in some few simple cases. But in most cases, this precise dynamics is unknown, or the calculations are untractable. Therefore $q$ can be obtained by fitting results with the analytic forms ($q$-exponentials, $q$-Gaussians) that emerge within nonextensive statistical mechanics. A third class of fitting parameters would be those which emerge from no specific theory, but are present in some convenient heuristic expressions. Within the $q$-statistical theory, we are not especially interested in those. \\

(vi) Dauxois considers {\it ``particularly pressing"} an undoubtedly relevant question, namely {\it ``does the $q$-Gaussian law describe the details of some physical problems?"}. Many systems exist that have been shown to exhibit, at the available precision, $q$-Gaussians. A particularly instructive example concerns cold atoms in dissipative optical lattices. It was predicted in 2003 by Lutz \cite{Lutz2003} that these atoms should have a $q$-Gaussian distribution of velocities with 
\begin{equation}
q=1+ \frac{44 E_R}{U_0} \,,
\end{equation}   
$E_R$ and $U_0$ being respectively the microscopic {\it recoil energy} and {\it potential depth}. In 2006, an impressive confirmation was provided \cite{DouglasBergaminiRenzoni2006}, through both quantum Monte Carlo simulations, and real experiments with $Cs$ atoms (the quality of the agreements are given by the correlation coefficients $R^2=0.995$ and $R^2=0.9985$ respectively).  Other examples (some of them briefly mentioned in \cite{Dauxois2007}) that, within reasonably good approximation, have exhibited $q$-Gaussians are the motion of cells of {\it Hydra viridissima} \cite{UpadhyayaRieuGlazierSawada2001}, defect turbulence \cite{DanielsBeckBodenschatz2004}, fluctuations of the magnetic field in the solar wind \cite{BurlagaVinas2004,VinasMaceBenson2005,BurlagaVinas2005,BurlagaNessAcuna2006,BurlagaVinasWang2007,BurlagaNessAcuna2007}, fluctuations of the temperature of the universe \cite{BernuiTsallisVillela2007}, financial return distributions \cite{Borland2002a,Borland2002b,Queiros2005a,Queiros2005b,QueirosMoyano2006,QueirosMoyanoSouzaTsallis2007,Queiros2007}, many-body Hamiltonian with long-range-interacting classical rotators approached through the HMF \cite{PluchinoRapisardaTsallis2007,TsallisRapisardaPluchinoBorges2007}, logistic map at edge of chaos \cite{TirnakliBeckTsallis2007}, granular matter \cite{ArevaloGarcimartinMaza2007}, earthquakes \cite{CarusoPluchinoLatoraVinciguerraRapisarda2007}, etc. \\   

(vii) {\it ``More importantly"}, as a second, {\it ``particularly pressing"} question, Dauxois asks {\it ``is anyone able to provide analytical predictions of the value of the $q$-index in terms of the microscopic parameters of the physical system?"}. This is of course an important point, that any quantitatively mature statistical mechanical theory (and physical theory in general) is expected to satisfy.  The Author might be unaware of the fact that many examples exist for which analytical calculations of $q$ have been possible from microscopic or mesoscopic dynamics. Let us mention some of them. Of course, Eq. (1) already constitutes one such example.  Another analytical result is the fact that the index $q$ has been connected \cite{Robledo2005} with the standard critical exponent $\delta$ (the exponent characterizing the order parameter as a function of its thermodynamically conjugate variable, at a second-order critical phenomenon) as follows:
\begin{equation}
q=\frac{1+\delta}{2} \,.
\end{equation}
In the area of quantum entanglement, the block entropy of fully entangled quantum magnetic chains involving only short-range interactions, has been recently proved to satisfy \cite{CarusoTsallis2007}
\begin{equation}
q=\frac{\sqrt{9+c^2}-3}{c} \,,
\end{equation}
where $c$ is the {\it central charge} emerging within conformal quantum field theory. This means that, for the Ising ferromagnet (i.e., $c=1/2$) we have $q=\sqrt{37}-6 = 0.08...$, and for the isotropic $XY$ ferromagnet (i.e., $c=1$), we have $q=\sqrt{10}-3 = 0.16...$. In the limit $c \to\infty$ the BG result $q=1$ is recovered. As another analytical connection, we may mention the Albert-Barabasi model in \cite{AlbertBarabasi2000} for (asymptotically) scale-free networks, for which we have (see \cite{ThurnerKyriakopoulosTsallis2007} for the connection of the Albert-Barabasi exponent and $q$)
\begin{equation}
q=\frac{2m(2-r)+1-p-r}{m(3-2r)+1-p-r} \,,
\end{equation}
where  $(m,p,r)$ are microscopic parameters of the model. Let us also mention that, on the grounds of \cite{TsallisBukman1996}, the following connection is since long predicted between $q$ and the anomalous diffusion space-time scaling exponent $\gamma$:
\begin{equation}
\gamma=\frac{2}{3-q} \,.
\end{equation}
This prediction has been verified by now in various systems, such as the {\it Hydra viridissima} cells \cite{UpadhyayaRieuGlazierSawada2001}, defect turbulence \cite{DanielsBeckBodenschatz2004}, $\alpha$-XY inertial ferromagnetic model \cite{RapisardaPluchino2005} (see \cite{AnteneodoTsallis1998} for details on the model), and granular matter flowing out from a silo \cite{ArevaloGarcimartinMaza2007}.

As our last present illustration of analytical predictions involving $q$, we may mention one which is available since 1999 \cite{CoradduKaniadakisLavagnoLissiaMezzoraniQuarati1999,CoradduLissiaMezzoraniQuarati2002,LissiaQuarati2005} and which has been successfully applied  to plasma astrophysics (e.g., solar plasma). 
For weakly nonideal plasma, we have \cite{CoradduKaniadakisLavagnoLissiaMezzoraniQuarati1999,CoradduLissiaMezzoraniQuarati2002,LissiaQuarati2005}
\begin{equation} 
|\,q-1| = 24 \, \alpha^4 \, \gamma^2 < 1 \,,
\end{equation}
where $\alpha$ (typically $0.4 < \alpha < 0.9$) is the {\it ion-ion correlation-function parameter} (see details in \cite{YanIchimaru1986,Ichimaru1992}), and
$\gamma \equiv (Ze)^2 n^{1/3} / k_BT$ is the {\it plasma parameter} (see Eq. (7) of \cite{CoradduKaniadakisLavagnoLissiaMezzoraniQuarati1999}), $n$ being the {\it average density}.   \\

(viii) The Author of \cite{Dauxois2007} makes, a few lines later, a short-cut which might mislead the quick reader. He states that {\it ``Hilhorst and Schehr show that $q$-Gaussians do not pass a careful inspection."}. Instead, what these Authors have (interestingly) shown is that the {\it MTG} and {\it TMNT} models do not {\it pass} in the very specific sense that their limiting distributions are {\it not exactly} but only approximatively $q$-Gaussians. This point was already discussed in the beginning of the present manuscript. \\

(ix) The Author of \cite{Dauxois2007} states that {\it ``if there is one lesson that has to be learned here, it is that one should be extremely careful when interpreting non-Gaussian data in terms of $q$-Gaussians"}. This is certainly true. But, of course, this also is true for Gaussians, Lorentzians, circles, ellipses, straight lines (in fact, $q$-Gaussians are mere straight lines in $\ln_q y \equiv (y^{1-q}-1)/(1-q)$ versus $x^2$ representation), parabolas, etc, all the analytical tools with which physicists have attempted to interpret and quantitatively understand the plethora of experimental, observational and computational results they are dealing with since centuries and millenia. But of course, --- {\it ce qui est clair sans le dire est encore plus clair en le disant!} ---, {\it any} experimental, observational or computational setup can only provide a {\it finite} amount of numbers with {\it finite} precision. This will {\it always} allow for an {\it infinite} number of analytic curves or  interpretations. The content of \cite{HilhorstSchehr2007} provides an interesting, and possibly fruitful, illustration of this obvious fact.\\

\section*{Acknowledgements}
Financial support by  CNPq and Faperj (Brazilian Agencies) is gratefully acknowledged. \\

\end{document}